\documentstyle[multicol,eqsecnum,aps,psfig]{revtex}
\renewcommand{\narrowtext}{\begin{multicols}{2}
\global\columnwidth20.5pc\noindent}
\renewcommand{\widetext}{\end{multicols}
\global\columnwidth42.5pc}
\begin{document}
\draft
\preprint{15 August 2001}
\title{Soliton excitations in halogen-bridged mixed-valence
       binuclear metal complexes}
\author{Shoji Yamamoto and Masanori Ichioka}
\address{Department of Physics, Okayama University,
         Tsushima, Okayama 700-8530, Japan}
\date{Received 15 August 2001}
\maketitle
\begin{abstract}
Motivated by recent stimulative observations in halogen
(X)-bridged binuclear transition-metal (M) complexes, which are
referred to as MMX chains, we study solitons in a one-dimensional
three-quarter-filled charge-density-wave system with both
intrasite and intersite electron-lattice couplings.
Two distinct ground states of MMX chains are reproduced and the
soliton excitations on them are compared.
In the weak-coupling region, all the solitons are degenerate to each
other and are uniquely scaled by the band gap, whereas in the
strong-coupling region, they behave differently deviating from the
scenario in the continuum limit.
The soliton masses are calculated and compared with those for
conventional mononuclear MX chains.
\end{abstract}
\narrowtext

\section{Introduction}\label{S:I}

Quasi-one-dimensional halogen (X)-bridged transition-metal (M)
complexes \cite{C95}, which are referred to as MX chains, have been
attracting much interest because of their unique behavior featured by
electron-electron correlations, electron-lattice interactions,
low dimensionality and $d$-$p$ orbital hybridization
\cite{G6408,W6435}.
The PtX compounds \cite{R217,D191,K357} have mixed-valence ground
states exhibiting intense and dichroic charge-transfer absorption,
strong resonance enhancement of Raman spectra and luminescence with
large Stokes shift.
On the other hand, the NiX compounds \cite{T4261,T2341,H8438} possess
monovalence magnetic ground states due to the strong on-site $d$-$d$
Coulomb interaction.
Substituting the bridging halogens, ligand molecules and counter ions
as well as the transition metals, we can widely tune the electronic
states of MX chains \cite{H9}.

   Solitonic excitations inherent in doubly degenerate
charge-density-wave systems stimulate further interest in MX
materials.
In this context, we may first be reminded of polyacetylene, the
trans isomer of which exhibits topological solitons, or moving domain
walls.
Electron-spin-resonance measurements \cite{S1565,G1132} on
trans-polyacetylenes demonstrated the existence of highly mobile
neutral magnetic defects, while infrared absorption spectra
\cite{F4140} for lightly doped ones illuminated the formation of
charged domain walls.
Such observations were finely interpreted employing a simple but
relevant model Hamiltonian \cite{S1698,S2099,T2388}.
The soliton picture established for polyacetylene led to an idea
\cite{I137,O250} of similar defect states existing in MX chains.
Although MX chains could not electrochemically be doped,
photogenerated solitons were extensively observed
\cite{K2122,O2248} in PtX chains.

   In recent years, a new class of halogen-bridged metal complexes,
which consist of alternating binuclear metal complexes and halogen
ions and are thus referred to as MMX chains, have renewed our
interest in this system.
In comparison with MX chains, MMX chains indeed look more
interesting.
The formal oxidation state of the metal ions is $3+$ in MX chains,
whereas it is $2.5+$ in MMX chains.
Therefore, MMX chains have an unpaired electron per metal dimer even
in their trapped-valence states, contrasting with the valence-trapped
state consisting of M$^{2+}$ and M$^{4+}$ in MX chains.
The M$(d_{z^2})$-M$(d_{z^2})$ direct overlap in MMX chains
effectively reduces the on-site Coulomb repulsion due to its
$d_{\sigma^*}$ character and therefore enhances the electrical
conductivity.

\begin{figure}
\begin{flushleft}
\mbox{\psfig{figure=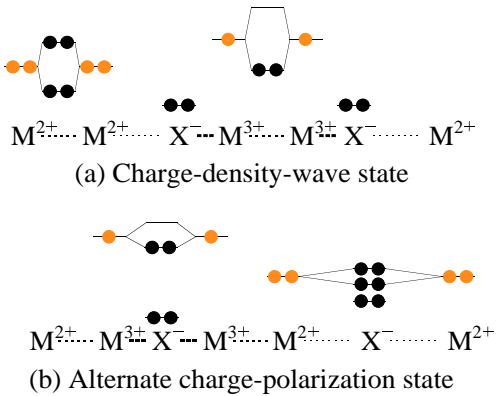,width=100mm,angle=0}}
\end{flushleft}
\caption{Schematic representation of the two distinct ground states
         of MMX chains:
         (a) Charge-density-wave (CDW) state whose X sublattice is
         dimerized;
         (b) Alternate charge-polarization (ACP) state whose M$_2$
         sublattice is dimerized.}
\label{F:CDW}
\end{figure}

   The actual observations of MMX chains are really stimulative.
The thus-far synthesized MMX compounds comprise two families:
 M$_2$(dta)$_4$I
($\mbox{M}=\mbox{Pt},\mbox{Ni}$;
 $\mbox{dta}=\mbox{dithioacetate}=\mbox{CH}_3\mbox{CS}_2^{\,-}$)
 \cite{B444,B2815}
 and
 R$_4$[Pt$_2$(pop)$_4$X]$\cdot$$n$H$_2$O
($\mbox{X}=\mbox{Cl},\mbox{Br},\mbox{I}$;
 $\mbox{R}=\mbox{Li},\mbox{K},\mbox{Cs},\cdots$;
 $\mbox{pop}=\mbox{diphosphonate}
 =\mbox{P}_2\mbox{O}_5\mbox{H}_2^{\,2-}$)
 \cite{C4604,C409}.
The former, dta complexes, are not yet well investigated but one of
them, Pt$_2$(dta)$_4$I, has now come into hot argument due to its
fascinating features \cite{K10068}.
Pt$_2$(dta)$_4$I exhibits metallic conduction around room
temperature, which is the first observation in one-dimensional
halogen-bridged metal complexes.
With decreasing temperature, there occurs a metal-semiconductor
transition at $300$ K and further transition to the
Peierls-insulating charge-ordering mode, which is shown in Fig.
\ref{F:CDW}(b), follows around $80$ K.
The alternate-charge-polarization (ACP) state is accompanied by
metal-sublattice dimerization, which has never been observed in MX
compounds.
Pt$_2$(dta)$_4$I has a neutral chain structure, where the metal
sublattice gets rid of any hydrogen-bond network and therefore
exhibits more pronounced one dimensionality.
The novel two-step thermal behavior has recently been supported
theoretically \cite{Y1198}.
On the other hand, the latter, pop complexes, have extensively been
measured \cite{C409,K4420,B1155,K40,W1195} and their ground states
have in general been assigned to charge-density-wave (CDW) states
of the conventional type, which are shown in Fig. \ref{F:CDW}(a).
However, a large choice of bridging halogens, ligand molecules and
counter ions in the pop complexes results in a wide tunability of
their ground states \cite{Y2321}.
Such a tuning of the electronic state can be realized also by
pressure \cite{S1405,S66,G1191}.
It is interesting that an applied pressure diminishes the Peierls
gap at low temperatures \cite{B239,B339} but oppositely stabilizes
the CDW state at higher temperatures \cite{Y}.

   Local states of MMX chains, whether photogenerated or
doping-induced ones, must be a coming potential subject.
The photoinduced-absorption spectrum has just been measured to reveal
the intrinsic charge-transfer excitations, but the photoexperiments
\cite{M} are still in their early stage.
Though the optical conductivity has recently been calculated
\cite{K}, any microscopic information on the defect states remains
lacking.
In such circumstances, we give a detailed report on soliton
excitations in MMX chains making full use of electron-phonon models.
Such an approach as regards electron-lattice interactions as most
important not only successfully interpreted the excitation spectrum
of polyacetylene \cite{S1698,S2099,T2388,K4173,C6862,O2478,N97} but
also clarified characteristic excitations \cite{O250,B339} of MX
chains.
The analytic aspect \cite{T2388,O250} of the class of electron-phonon
models further stimulates our interest.
Since recent extensive calculations \cite{Y1198,Y125124} of the MMX
system have shown that the dta and pop complexes are mainly
characterized by their distinct electron-lattice interactions, our
starting point is well justified from the practical point of view as
well.
Let us take the first step toward explorations into nonlinear
excitations in mixed-valence binuclear metal complexes.
\vskip -30mm

\section{Ground-State Properties}\label{S:GSP}

   We introduce the $\frac{3}{4}$-filled one-dimensional single-band
two-orbital electron-phonon model:
\begin{eqnarray}
   {\cal H}
    &=&-t_{\rm MM}\sum_{n,s}
        (a_{n,s}^\dagger b_{n,s}+b_{n,s}^\dagger a_{n,s})
     \nonumber \\
     &-&\sum_{n,s}
        \bigl[
         t_{\rm MXM}-\alpha(v_{n+1}-v_n)
        \bigr]
        (b_{n,s}^\dagger a_{n+1,s}+a_{n+1,s}^\dagger b_{n,s})
     \nonumber \\
     &-&\beta\sum_{n,s}
        \bigl[
         (v_n-u_{n-1})a_{n,s}^\dagger a_{n,s}
        +(u_n-v_n    )b_{n,s}^\dagger b_{n,s}
        \bigr]
     \nonumber \\
     &+&\frac{K_{\rm MX}}{2}\sum_{n}
        \bigl[
         (u_n-v_n)^2+(v_n-u_{n-1})^2
        \bigr]\,,
   \label{E:H}
\end{eqnarray}
where $a_{n,s}^\dagger$ and $b_{n,s}^\dagger$ are the creation
operators of an electron with spin $s=\pm$ (up and down) for the M
$d_{z^2}$ orbitals in the $n$th MMX unit, which we refer to as $a$-
and $b$-metal sites.
$t_{\rm MM}$ and $t_{\rm MXM}$ describe the  electron hoppings
between neighboring metal sites.
$\alpha$ and $\beta$ are the intersite and intrasite electron-phonon
coupling constants, respectively, with $K_{\rm MX}$ being the
metal-halogen spring constant.
$u_n$ and $v_n$ are, respectively, the chain-direction displacements
of the halogen and metal dimer in the $n$th MMX unit from their
equilibrium position.
We assume, based on the thus-far reported experimental observations,
that every M$_2$ moiety is not deformed.
We set $t_{\rm MM}$ and $K$ both equal to unity unless a particular
mention.
\begin{figure}
\centerline
{\mbox{\psfig{figure=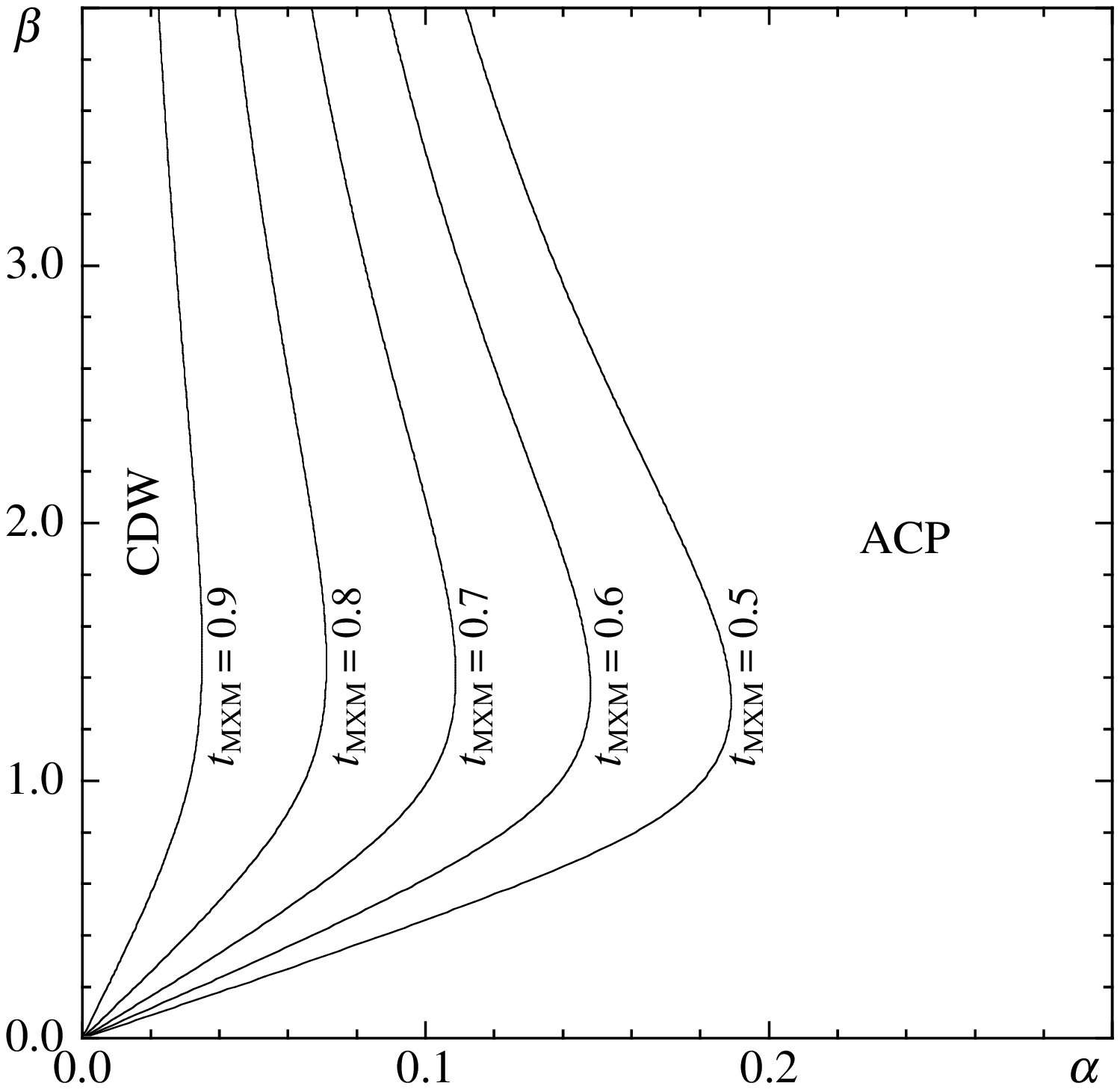,width=58mm,angle=0}}}
\vskip -30mm
\caption{The $\alpha$-versus-$\beta$ ground-state phase diagram as a
         function of $t_{\rm MXM}$.}
\label{F:PhD}
\vskip -27mm
\centerline
{\qquad\qquad\mbox{\psfig{figure=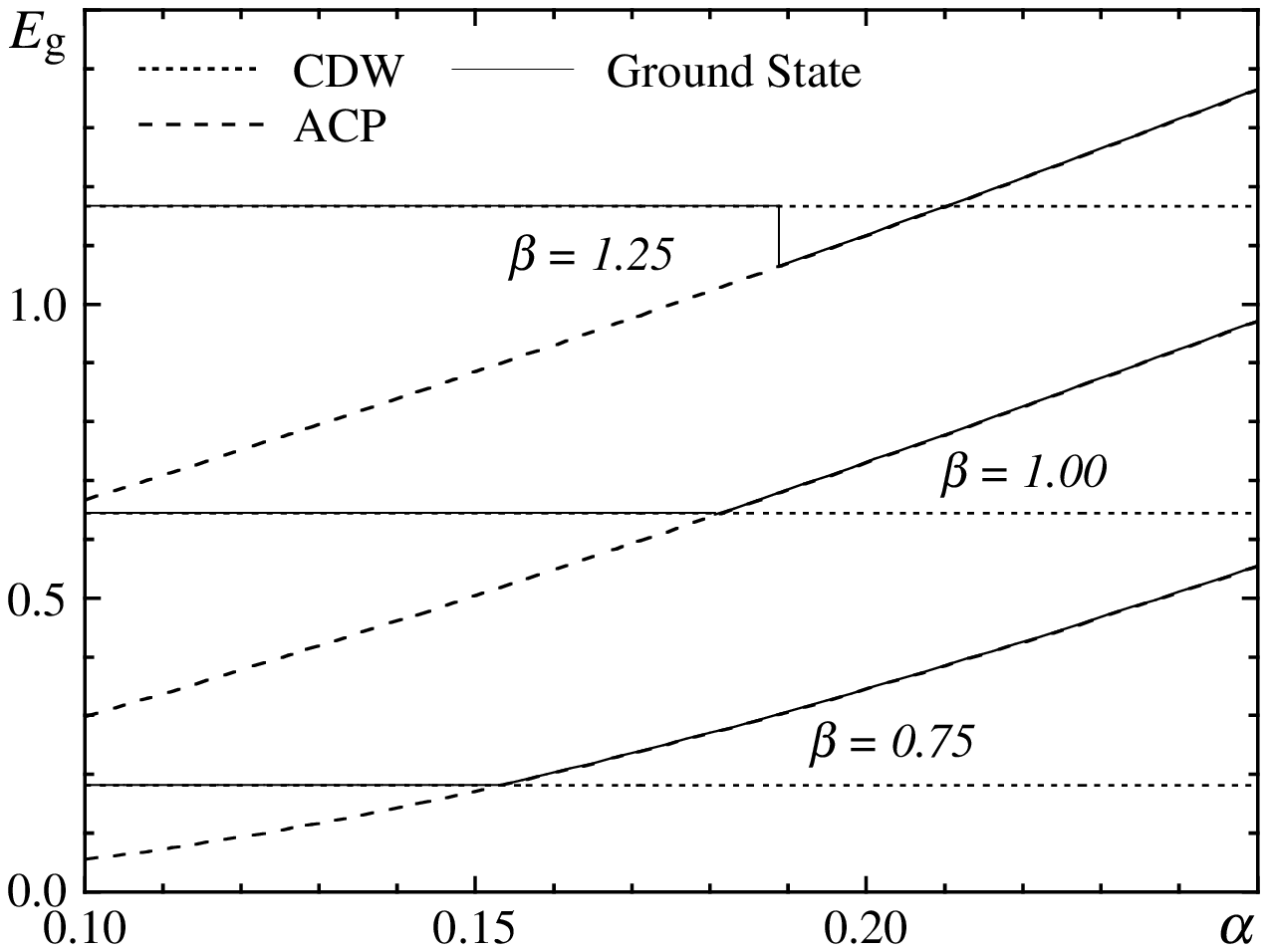,width=83mm,angle=0}}}
\vskip 3mm
\caption{The competing Peierls gaps of the CDW and ACP states and the
         consequent band gap as functions of $\alpha$ and $\beta$.}
\label{F:Eg}
\end{figure}
\vskip -4mm

   This Hamiltonian possesses two distinct ground states which are
shown in Fig. \ref{F:CDW} and their competition is visualized in Fig.
\ref{F:PhD}.
The CDW state is characterized by the alternating on-site energies,
whereas the ACP state comparatively by the alternating interdimer
transfer energies.
The orbital hybridization  within every M$_2$ moiety is essential in
the valence-trapped CDW state, while it is the overlap of the
$d_{\sigma^*}$ orbitals on neighboring M$_2$ moieties that stabilizes
the valence-delocalized ACP state.
Therefore, increasing $\beta$ and $t_{\rm MM}$ advantageously act on
the CDW state, whereas increasing $\alpha$ and $t_{\rm MXM}$ on the
ACP state.
When $t_{\rm MM}=t_{\rm MXM}$ and $\alpha=0$, the CDW and ACP states
are degenerate to each other regardless of $\beta$.

   The phase transition between the two states is of the first order.
We plot in Fig. \ref{F:Eg} the Peierls gaps as functions of the
coupling constants.
The band gap varies continuously across the phase boundary in the
weak-coupling region, whereas it exhibits a discontinuity in the
strong-coupling region.
The ACP state depends on both $\alpha$ and $\beta$, but the CDW state
has no dependence on $\alpha$.
The intersite electron-phonon coupling originates in distortion of
the M$_2$ sublattice.

   The calculations are well consistent with the experimental
observations; the pop-family compounds (NH$_4$)$_4$[Pt$_2$(pop)$_4$X]
exhibit ground states of the CDW type \cite{K40}, while the dta
complex Pt$_2$(dta)$_4$I displays that of the ACP type \cite{K10068}.
The M$_2$ moieties are tightly locked together in the pop complexes
due to the hydrogen bonds between the ligands and the counter
cations, whereas they are much more movable in Pt$_2$(dta)$_4$I owing
to its neutral chain structure.
Thus, considering that $\alpha$ indirectly describes the mobility of
the M$_2$ sublattice, a significantly larger $\alpha$ is expected for
Pt$_2$(dta)$_4$I.
\vskip 1mm
\begin{figure}
\centerline
{\qquad\quad\mbox{\psfig{figure=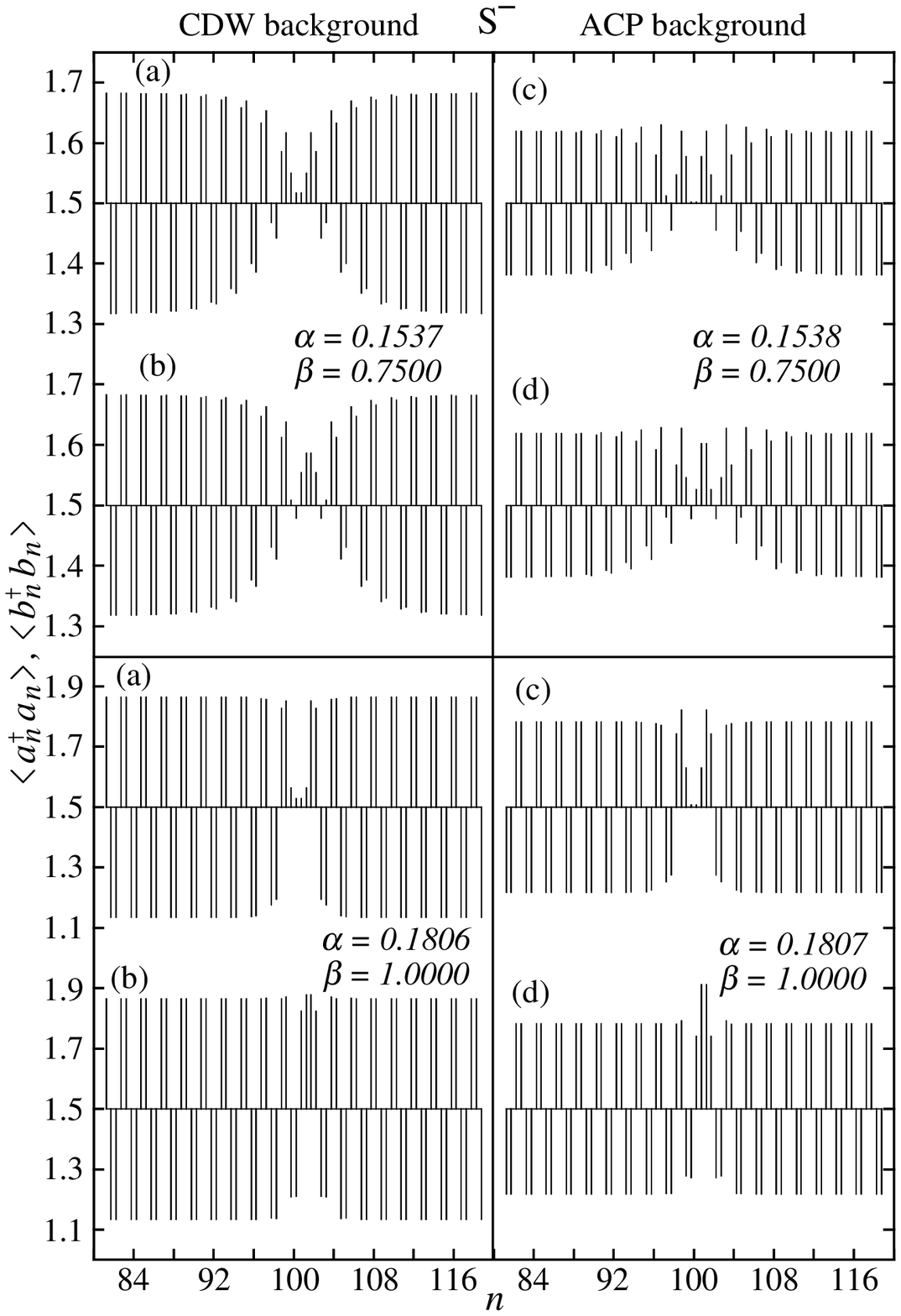,width=82mm,angle=0}}}
\vskip 2mm
\caption{Electronic structures of the negatively charged solitons in
         the CDW (a and b) and ACP (c and d) states, where quantum
         averages of the local electron densities are measured in
         comparison with the average occupancy:
         (a) $x_0=100.5$ with the highest energy;
         (b) $x_0=101.5$ with the lowest energy;
         (c) $x_0=100$ with the highest energy;
         (d) $x_0=101$ with the lowest energy.}
\label{F:S-}
\centerline
{\qquad\quad\mbox{\psfig{figure=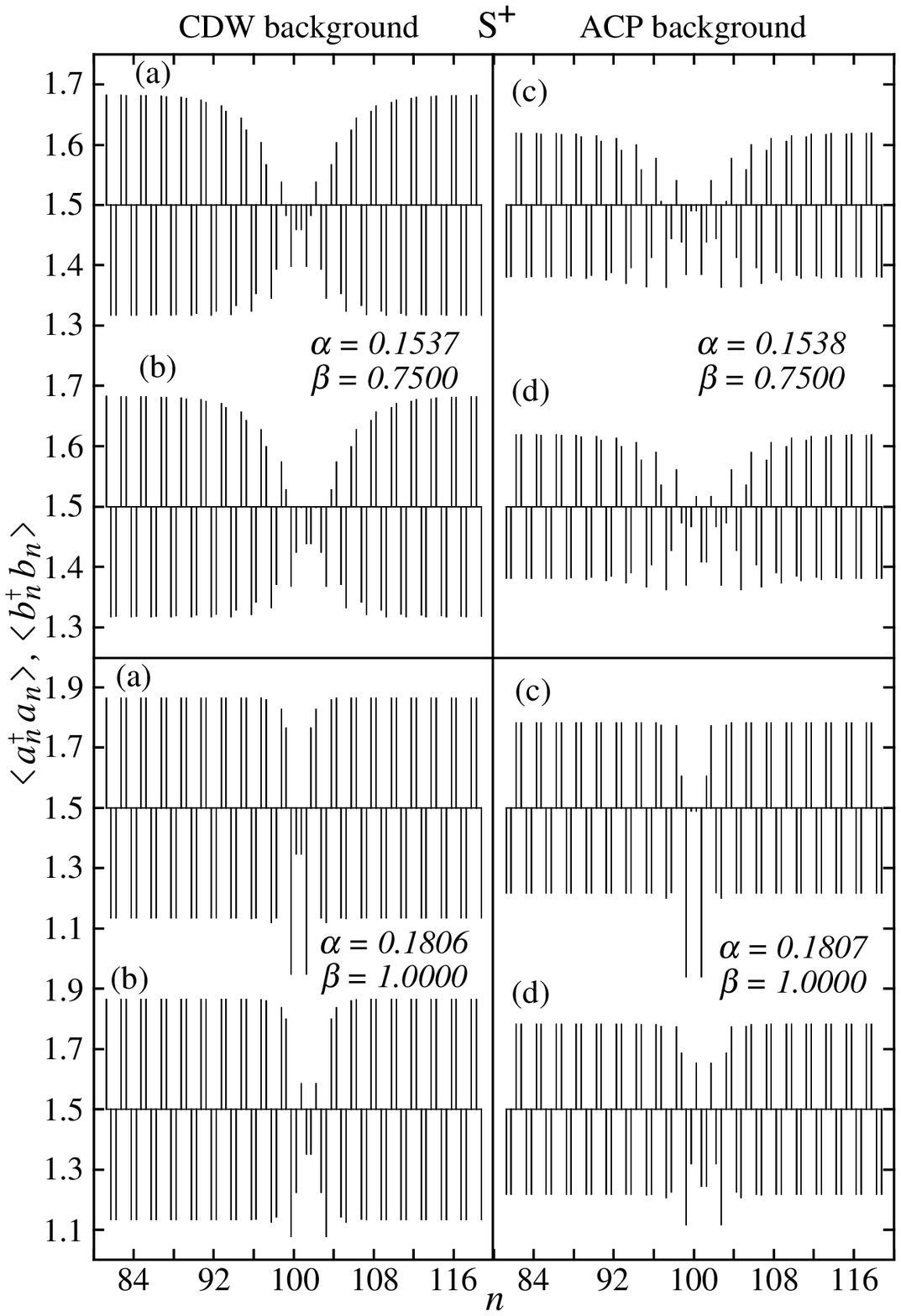,width=82mm,angle=0}}}
\vskip 2mm
\caption{The same as Fig. 4 but the positively charged solitons:
         (a) $x_0=100.5$ with the lowest energy;
         (b) $x_0=101.5$ with the highest energy;
         (c) $x_0=100$ with the lowest energy;
         (d) $x_0=101$ with the highest energy.}
\label{F:S+}
\vskip 3mm
\centerline
{\qquad\quad\mbox{\psfig{figure=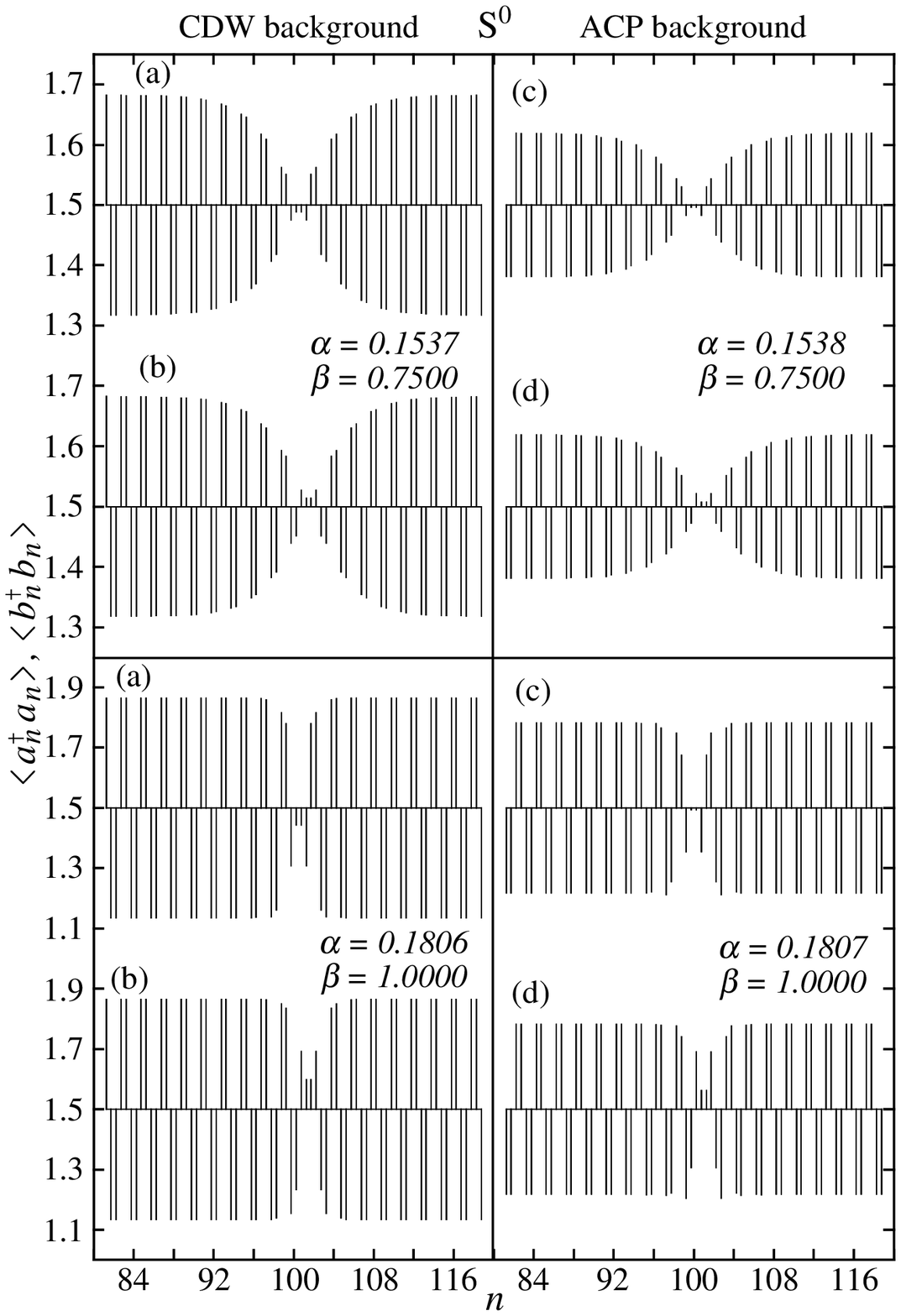,width=82mm,angle=0}}}
\vskip 2mm
\caption{The same as Fig. 4 but the neutral solitons:
         (a) $x_0=100.5$ with the highest energy;
         (b) $x_0=101.5$ with the lowest energy;
         (c) $x_0=100$ with the highest energy;
         (d) $x_0=101$ with the lowest energy.}
\label{F:S0}
\end{figure}

\section{Soliton Excitations}\label{S:SE}

   Now we search for soliton solutions of the Hamiltonian
(\ref{E:H}).
Under the constraint of the total chain length being unchanged, a
trial wave function may be introduced as
\cite{S1698,S2099}
\begin{equation}
   u_n-v_n
  =\sigma(v_{n+1}-u_n)
  =(-1)^n l_0{\rm tanh}
   \bigl[
    (na-x_0)/\xi_{\rm s}
   \bigr]\,,
   \label{E:WF}
\end{equation}
where $\sigma$ takes $-$ and $+$ for the CDW and ACP backgrounds,
respectively, $a$ is the lattice constant (the original unit-cell
length), $l_0$ is the metal-halogen bond-length change in the ground
state, and $x_0$ and $\xi_{\rm s}$ are, respectively, the
soliton center and width, both of which are variationally determined.
Since we impose the periodic boundary condition on the Hamiltonian,
the soliton solutions demand that the number of the original unit
cells, $N$, should be odd.
We set $N$ equal to $201$, which is much larger than $\xi_{\rm s}$
in any calculation of ours.
Then, negatively-charged (S$^-$), positively-charged (S$^+$) and
neutral (S$^0$) solitons are obtained by setting $N_+=N_-=302$,
$N_+=N_-=301$ and $N_\pm=N_\mp+1=302$, respectively, where $N_s$ is
the number of the electrons with spin $s$.
Due to the spin-rotational symmetry, the neutral solitons with up and
down spins are degenerate to each other.
The degeneracy between the neutral and charged solitons may be lifted
in the strong-coupling region.
When we compare the solitons with the CDW and ACP backgrounds, we
calculate them in the vicinity of the phase boundary so as to
illuminate their essential differences, if any.
Taking the structural analyses \cite{B444,C4604,C409} into
consideration, we set $t_{\rm MM}=2t_{\rm MXM}$.

\subsection{Wave Functions}\label{SS:WF}

   We show the spatial configurations of S$^-$, S$^+$ and S$^0$ in
Figs. \ref{F:S-}, \ref{F:S+} and \ref{F:S0}, respectively.
Their formation energies $E_{\rm s}$ do not depend on their locations
in the weak-coupling region, but the degeneracy is lifted in the
strong-coupling region.
Two of the differently-located solitons are shown at each
parametrization, one of which has the highest energy as a function
of $x_0$, while the other of which is the optimum configuration.
As the coupling strength increases, solitons generally possess
increasing energies and end up with immobile defects.
\vskip 2mm

\begin{figure}
\centerline
{\qquad\quad\mbox{\psfig{figure=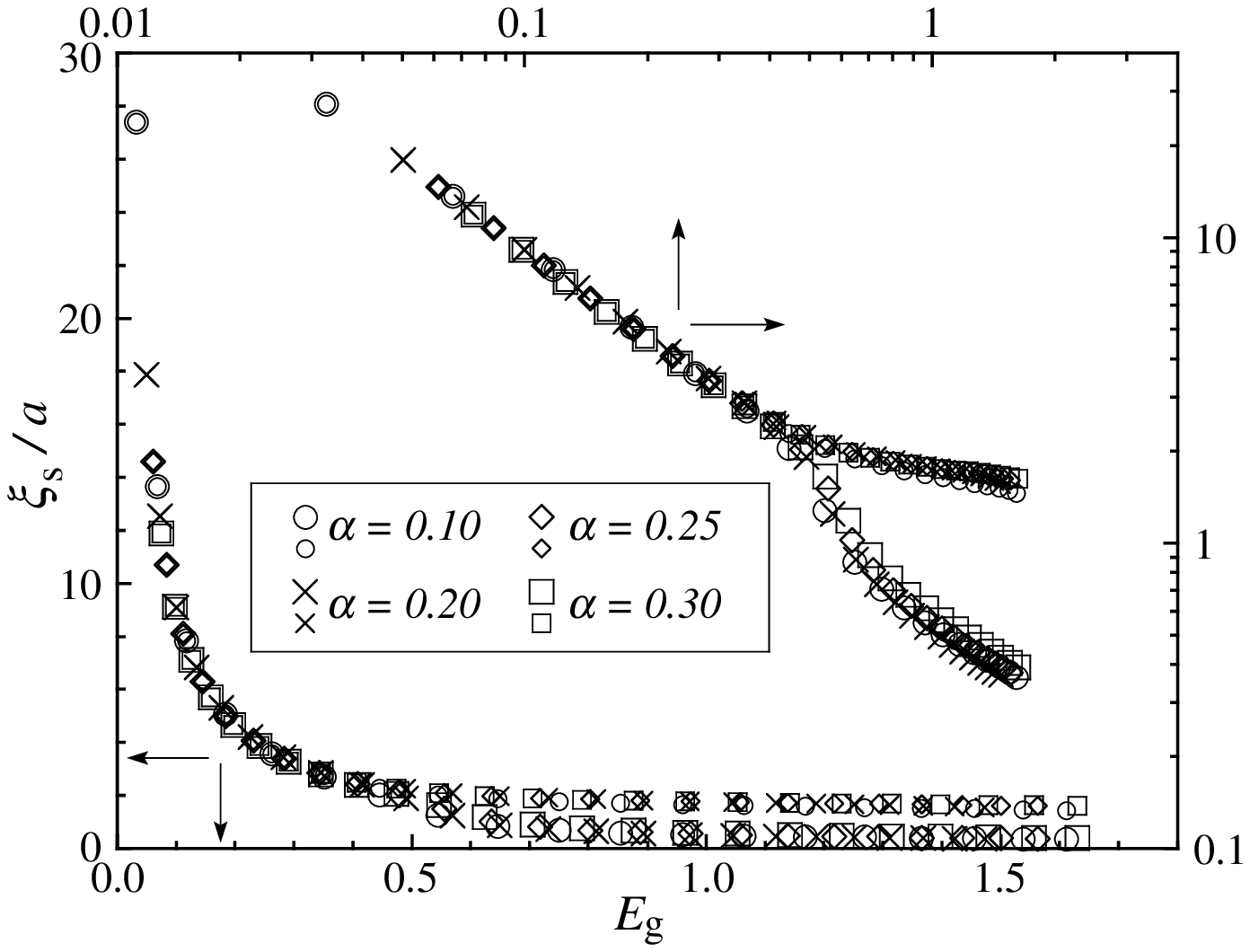,width=92mm,angle=0}}}
\vskip 2mm
\caption{The optimized width of the negatively charged soliton as a
         function of the band gap $E_{\rm g}$ in linear and
         logarithmic scales under various parametrizations, where the
         larger and smaller symbols correspond to the lowest- and
         highest-energy locations $x_0$, respectively.}
\label{F:xi}
\centerline
{\qquad\qquad\mbox{\psfig{figure=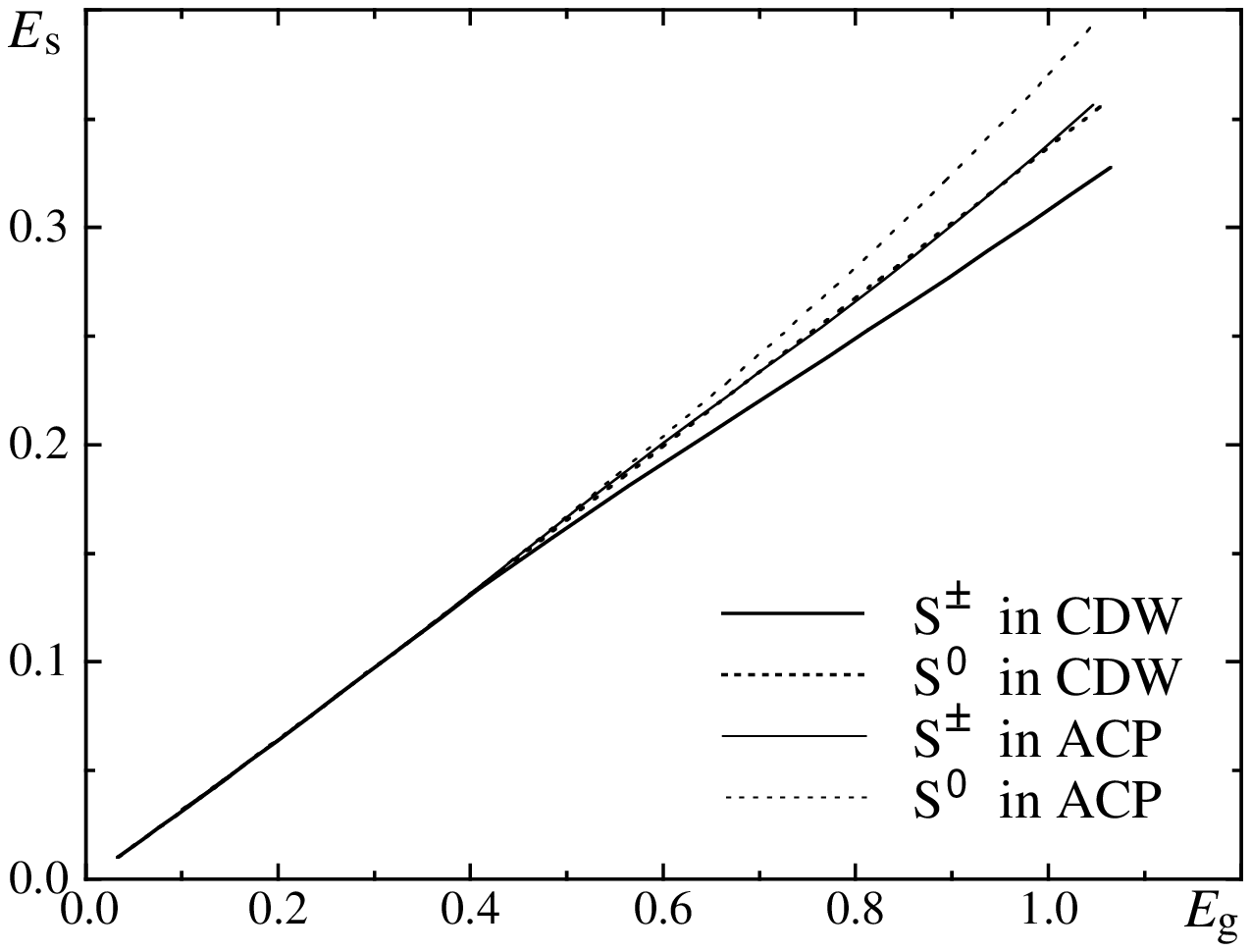,width=86mm,angle=0}}}
\vskip 3mm
\caption{The soliton formation energies $E_{\rm s}$ as functions of
         the band gap $E_{\rm g}$, where $E_{\rm s}$ is averaged over
         $x_0$.
         The negatively and positively charged solitons possess more
         and less than three-quarter-filled electron bands,
         respectively, and therefore their formation energies can not
         be defined in themselves.
         As for the charged solitons, $E_{\rm s}$ is further averaged
         over S$^+$ and S$^-$.}
\label{F:Es}
\end{figure}

   $\xi_{\rm s}$ looks like a decreasing function of $\alpha$ and
$\beta$.
In order to illuminate its scaling behavior in more detail, we plot
in Fig. \ref{F:xi} $\xi_{\rm s}$ as a function of the band gap
$E_{\rm g}$.
Although we take $\alpha$ and $\beta$ at random, $\xi_{\rm s}$ is
uniquely scaled by $E_{\rm g}$ as far as $E_{\rm g}$ stays not so
large.
We obtain a scaling formula
\begin{equation}
   \xi_{\rm s}/a\simeq 0.96/E_{\rm g}^{0.98}\,,
   \label{E:xi}
\end{equation}
though there is large uncertainty in the second decimal place
according to data set adopted into the fitting.
The scaling relation fits all the solitons S$^\pm$ and S$^0$.
It is quite convincing that such a scaling law breaks down as
$\xi_{\rm s}$ approaches $a$, where the chain no more behaves as a
continuum.
We return to Eq. (\ref{E:xi}) later in an argument in the continuum
limit.

   The degeneracy with respect to the soliton type and location is
lifted in the strong-coupling region where $\xi_{\rm s}\alt a$.
In Fig. \ref{F:Es} we plot $E_{\rm s}$ as a function of $E_{\rm g}$
for various soliton solutions.
As the band gap grows, S$^0$ generally comes to have a higher energy
than S$^\pm$, which is the case in MX chains as well.
The simplest argument in the strong-coupling region may be the atomic
treatment of electrons.
Setting $t_{\rm MXM}=0$ in an MX chain with the CDW ground state, we
obtain the soliton energies as $E_{\rm s}(\mbox{S}^0)=3E_{\rm g}/4$
and $E_{\rm s}(\mbox{S}^\pm)=E_{\rm g}/2$ \cite{B339}, where
$2E_{\rm g}\equiv 4\beta^2/K$ is the Peierls gap in the atomic limit.
A similar argument in the present case is not so trivial.
The CDW and ACP states are stabilized by sufficient M-M intradimer
and M-X-M interdimer electronic communications, respectively
(see Fig. \ref{F:CDW}).
Therefore, even in the strong-coupling region, we can not reasonably
set $t_{\rm MM}=t_{\rm MXM}=0$ but should treat the hybridized
$d_{\sigma^*}$ orbitals under the situations,
$t_{\rm MXM}\ll t_{\rm MM}\alt\beta$ and
$t_{\rm MM}\ll t_{\rm MXM}\alt\beta$, for the CDW and ACP
backgrounds, respectively.
Such calculations still imply that
$E_{\rm s}(\mbox{S}^\pm)<E_{\rm s}(\mbox{S}^0)$.
\begin{figure}
\centerline
{\qquad\qquad\ \mbox{\psfig{figure=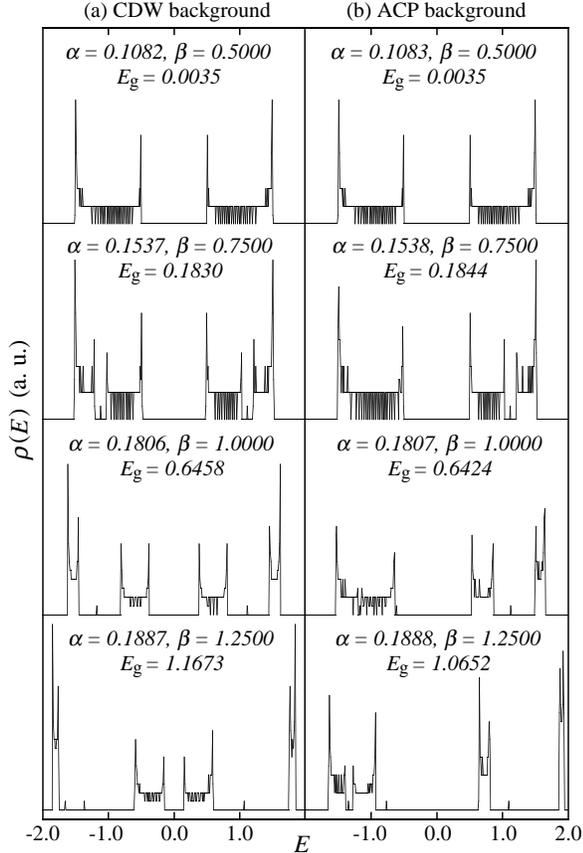,width=94mm,angle=0}}}
\vskip 1mm
\caption{Density of states $\rho(E)$ for the optimum soliton
         solutions with the CDW (a) and ACP (b) backgrounds.}
\label{F:DOS}
\end{figure}

\subsection{Energy Structures}\label{SS:ES}

   Density of states for the optimum soliton solutions are shown in
Fig. \ref{F:DOS}.
The background spectrum generally consists of four bands due to the
ground-state cell doubling.
The four bands of the CDW state are, from the bottom to the top,
largely made up of the bonding combination $\phi_+$ of binuclear
Pt$^{2+}$-Pt$^{2+}$ units, that of Pt$^{3+}$-Pt$^{3+}$ units, the
antibonding combination $\phi_-$ of Pt$^{2+}$-Pt$^{2+}$ units and
that of Pt$^{3+}$-Pt$^{3+}$ units.
On the other hand, the major components of the four bands of the ACP
state are
the $\phi_+$ orbitals of interdimer Pt$^{2+}$-X$^-$-Pt$^{2+}$ units,
the $\phi_-$ orbitals of Pt$^{2+}$-X$^-$-Pt$^{2+}$ units,
the $\phi_+$ orbitals of Pt$^{3+}$-X$^-$-Pt$^{3+}$ units and
the $\phi_-$ orbitals of Pt$^{3+}$-X$^-$-Pt$^{3+}$ units.
Thus, increasing $\beta$ splits both $\sigma$ and $\sigma^*$
orbitals in the CDW state but comparatively enhances the splitting
between $\sigma$ and $\sigma^*$ orbitals in the ACP state.

   The optimum soliton solutions commonly exhibit an additional level
within the gap.
In contrast with the case of polyacetylene \cite{S1698,S2099}, the
soliton levels generally deviate from the middle of the gap in MMX
chains as a result of the breakdown of the electron-hole symmetry.
Besides the intragap level, there appear a few related levels
depending upon the coupling strength.
As far as we work with electron-phonon Hamiltonians without any
Coulomb interactions, the level structure does not vary with the
soliton charge.

\begin{figure}
\centerline
{\qquad\quad\ 
 \mbox{\psfig{figure=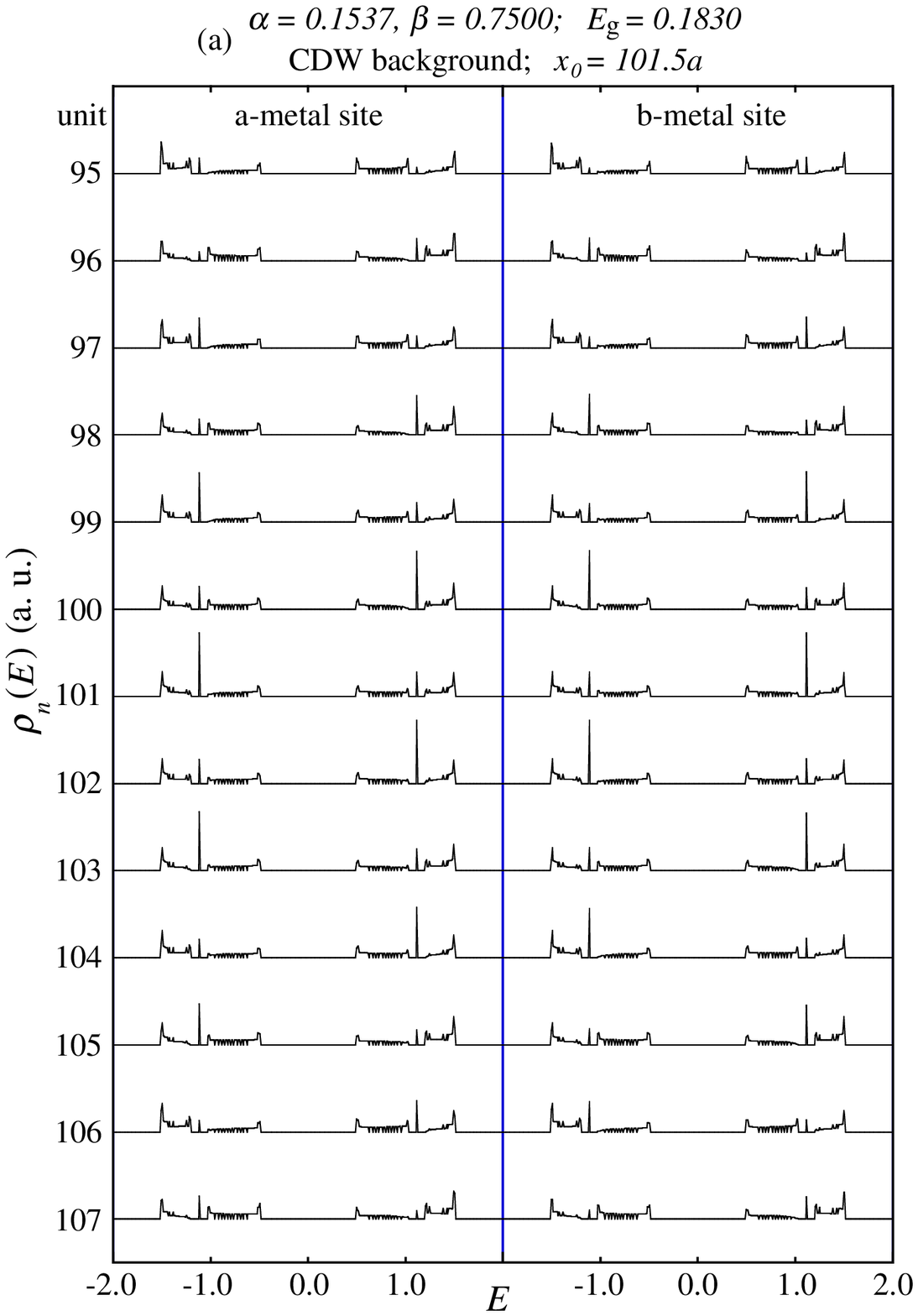,width=100mm,angle=0}
       $\!\!\!\!\!\!\!\!\!\!\!\!\!\!\!\!\!\!\!\!\!\!\!\!$
       \psfig{figure=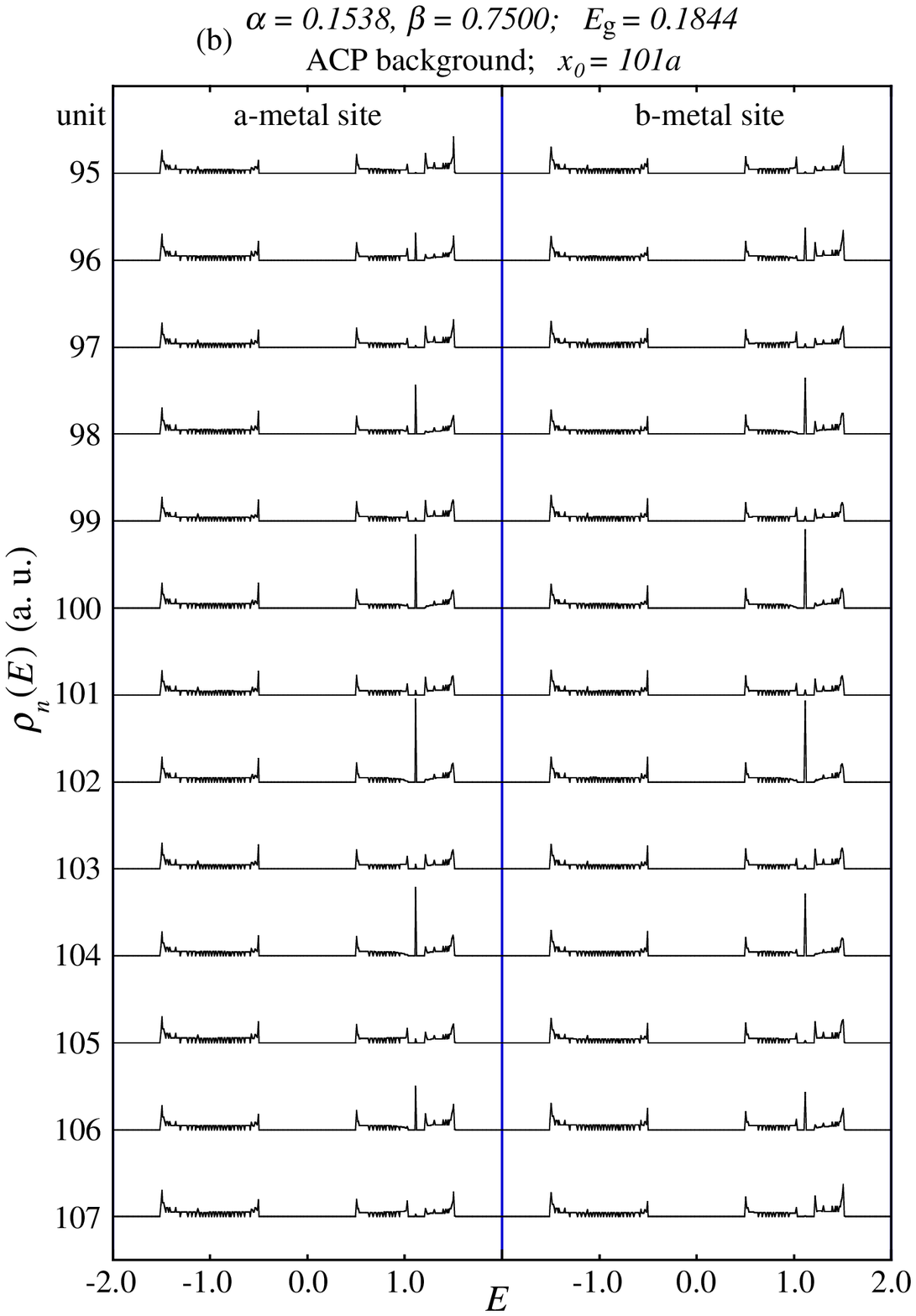,width=100mm,angle=0}}}
\vskip 1mm
\caption{Local density of states $\rho_n(E)$ for the optimum soliton
         solutions with the CDW (a) and ACP (b) backgrounds.}
\label{F:LDOS}
\end{figure}

   Figure \ref{F:LDOS} shows local density of states for the optimum
soliton solutions.
Solitons are indeed localized around their centers.
The probability density of the solitons generally oscillates
according as $n$ is even or odd, which is the same observations that
we find in the Su-Schrieffer-Heeger (SSH) model \cite{S1698,S2099} of
polyacetylene.
The oscillation is more remarkable on the ACP background, which is
convincing when we compare the binuclear metal units to the CH groups
in polyacetylene.
However, the ACP solitons in MMX chains have little probability at
their centers in contrast with the SSH solitons exhibiting the
largest probability at their centers.
Their opposite oscillating behaviors may be ascribed to their
distinct electron-phonon interactions of the Holstein (intrasite)
and SSH (intersite) type.

\section{Discussion}\label{S:D}

\subsection{The Continuum Limit}\label{SS:CL}

   The discovery of a rigorous soliton solution in the
Takayama-Lin-Liu-Maki (TLM) model \cite{T2388}, which is the
continuum version of the SSH model for polyacetylene, significantly
stimulated the theoretical investigations of MX chains.
On the analogy of the twofold degeneracy of the ground state,
Ichinose \cite{I137} proposed an idea of topological solitons
governing the lattice relaxation of charge-transfer excitations in
mixed-valence MX chains.
Onodera \cite{O250} further demonstrated that solitons should play an
important role in MX chains as well, pointing out the equivalence of
an MX chain to a CH chain in the continuum limit.
Since then nonlinear excitations in MX chains have also extensively
been calculated
\cite{G6408,W6435,B239,B339,C723,G10556,M5758,M5763,S1605,I1380,T4074,T2212}
and indeed been observed experimentally
\cite{K2122,O2248,K1789,H5706,D49}.

   Figure \ref{F:xi} has revealed that there still holds a scaling
law for soliton excitations in the MMX system.
In the weak-coupling region, the soliton width is uniquely scaled by
the band gap regardless of the ground-state properties.
The TLM model powerfully illuminated the scaling properties of CH
chains and was successfully tuned so as to describe MX chains.
Here we consider the continuum description of MMX chains.
\widetext
\narrowtext

   In order to derive the continuum version of the present model
(\ref{E:H}), it is essential to describe an MMX chain in terms of
effectively half-filled electron bands.
Therefore, we first neglect the electron-phonon coupling in the
Hamiltonian (\ref{E:H}) and obtain
\begin{equation}
   \lim_{\alpha,\beta\rightarrow 0}{\cal H}
   =\sum_{k,s}
    \left(
     \varepsilon_k^+ A_{k,s}^\dagger A_{k,s}
    +\varepsilon_k^- B_{k,s}^\dagger B_{k,s}
    \right)\,,
\end{equation}
where
\begin{equation}
   \varepsilon_k^\pm
   =\pm\sqrt{t_{\rm MM}^2+t_{\rm MXM}^2
            +2t_{\rm MM}t_{\rm MXM}{\rm cos}ka}\,,
\end{equation}
\begin{equation}
   \left.
   \begin{array}{lll}
    A_{k,s}
    &=&\frac{\displaystyle 1}{\displaystyle\sqrt{2}}
       \left(
        {\rm e}^{{\rm  i}\theta/2}a_{k,s}
       +{\rm e}^{{\rm -i}\theta/2}b_{k,s}
       \right)\,,\\
    B_{k,s}
    &=&\frac{\displaystyle 1}{\displaystyle\sqrt{2}}
       \left(
        {\rm e}^{ {\rm i}\theta/2}a_{k,s}
       -{\rm e}^{-{\rm i}\theta/2}b_{k,s}
       \right)\,,\\
   \end{array}
   \right.
\end{equation}
with
\begin{equation}
   {\rm e}^{{\rm i}\theta}
   =-\frac{t_{\rm MM}+t_{\rm MXM}{\rm e}^{{\rm i}ka}}
          {\sqrt{t_{\rm MM}^2+t_{\rm MXM}^2
                +2t_{\rm MM}t_{\rm MXM}{\rm cos}ka}}\,,
\end{equation}
\begin{equation}
   \left.
   \begin{array}{lll}
    a_{k,s}
    &=&\frac{\displaystyle 1}{\displaystyle\sqrt{N}}
       {\displaystyle\sum_{k,s}}{\rm e}^{-{\rm i}kna} a_{n,s}\,,\\
    b_{k,s}
    &=&\frac{\displaystyle 1}{\displaystyle\sqrt{N}}
       {\displaystyle\sum_{k,s}}{\rm e}^{-{\rm i}kna} b_{n,s}\,.\\
   \end{array}
   \right.
\end{equation}
Since we are interested in the low-lying excitations governed by the
electronic structure near the Fermi level, we discard the irrelevant
band $\varepsilon_k^-$ and linearize the relevant dispersion relation
$\varepsilon_k^+$ at the two Fermi points.
Then, assuming the CDW and ACP ground states, we reach effective
Hamiltonians in the coordinate representation:
\begin{eqnarray}
   &&
   {\cal H}_{\rm eff}^{\rm CDW}
  =t_{\rm eff}\sum_{n,s}
   \left[
    1-\frac{{\mit\Delta}(na)}{t_{\rm MM}}
   \right]
   \left(
    A_{n,s}^\dagger A_{n+1,s}+{\rm H.c.}
   \right)
   \nonumber \\
   &&\ 
  +\frac{2t_{\rm eff}}{t_{\rm MXM}}\sum_{n,s}
   {\mit\Delta}(na)
   A_{n,s}^\dagger A_{n,s}
  +\frac{K_{\rm MX}}{4\beta_{\rm eff}^2}\sum_n
   {\mit\Delta}(na)^2\,,
   \\
   &&
   {\cal H}_{\rm eff}^{\rm ACP}
  =t_{\rm eff}\sum_{n,s}
   \left[
    1-\frac{{\mit\Delta}(na)}{t_{\rm MXM}}
   \right]
   \left(
    A_{n,s}^\dagger A_{n+1,s}+{\rm H.c.}
   \right)
   \nonumber \\
   &&\ 
  +\frac{2t_{\rm eff}}{t_{\rm MM}}\sum_{n,s}
   {\mit\Delta}(na)
   A_{n,s}^\dagger A_{n,s}
  +\frac{K_{\rm MX}}{16\alpha_{\rm eff}^2}\sum_n
   {\mit\Delta}(na)^2\,,
\end{eqnarray}
where
\begin{equation}
   t_{\rm eff}=\frac{t_{\rm MM}t_{\rm MXM}}
                    {2\sqrt{t_{\rm MM}^2+t_{\rm MXM}^2}}\,,
\end{equation}
\begin{equation}
   \alpha_{\rm eff}
   =\frac{1}{2}
    \left(
     \alpha+\frac{t_{\rm eff}}{t_{\rm MM}}\beta
    \right)\,,\ \ 
   \beta_{\rm eff}
   =\frac{t_{\rm eff}}{t_{\rm MXM}}\,,
\end{equation}
\begin{equation}
   {\mit\Delta}(na)
   =\left\{
     \begin{array}{l}
      2\beta_{\rm eff}(-1)^n(u_n-v_n)\ \ \ 
      {\rm for}\ {\cal H}_{\rm eff}^{\rm CDW}\,,\\
      4\alpha_{\rm eff}(-1)^n(u_n-v_n)\ \ \ 
      {\rm for}\ {\cal H}_{\rm eff}^{\rm ACP}\,.\\
     \end{array}
    \right.
\end{equation}
The gap parameter ${\mit\Delta}(na)$ is constant in the ground state
and is expected to show smooth and slow spatial variations for the
relevant fluctuations.
Correspondingly we introduce slowly-varying field operators
\begin{equation}
   \left.
   \begin{array}{lll}
    \psi_s^{(r)}(na)
    &=&\frac{\displaystyle 1}{\displaystyle\sqrt{L}}
       {\displaystyle\sum_k}
       {\rm e}^{{\rm i}kna}A_{ k_{\rm F}+k,s}\,,\\
    \psi_s^{(l)}(na)
    &=&\frac{\displaystyle -{\rm i}}{\displaystyle\sqrt{L}}
       {\displaystyle\sum_k}
       {\rm e}^{{\rm i}kna}A_{-k_{\rm F}+k,s}\,,\\
   \end{array}
   \right.
\end{equation}
for right- and left-moving electrons, respectively, where
$L=Na$, $k_{\rm F}=\pi/2a$ and $-\pi/2a<k\leq\pi/2a$.
Now, summing out fast-varying components and taking the continuum
limit $a\rightarrow 0$, the lattice description can be mapped
onto a continuous line as
\begin{eqnarray}
   &&
   {\cal H}_{\rm eff}^{\rm CDW}
  =\int
   \frac{{\mit\Delta}(x)^2}{2\pi\hbar v_{\rm F}\lambda}
   {\rm d}x
  +\sum_s\int
   {\mit\Phi}_s^\dagger(x)
   \biggl[
    {\rm i}\hbar v_{\rm F}\sigma_z\frac{{\rm d}}{{\rm d}x}
   \nonumber \\
   &&\quad
  -2{\mit\Delta}(x)t_{\rm eff}
    \left(
     \frac{\sigma_x}{t_{\rm MM}}+\frac{\sigma_y}{t_{\rm MXM}}
    \right)
   \biggr]
   {\mit\Phi}_s(x)
   {\rm d}x\,,
   \label{E:HCDWc}
   \\
   &&
   {\cal H}_{\rm eff}^{\rm ACP}
  =\int
   \frac{{\mit\Delta}(x)^2}{2\pi\hbar v_{\rm F}\lambda}
   {\rm d}x
  +\sum_s\int
   {\mit\Phi}_s^\dagger(x)
   \biggl[
    {\rm i}\hbar v_{\rm F}\sigma_z\frac{{\rm d}}{{\rm d}x}
   \nonumber \\
   &&\quad
  -2{\mit\Delta}(x)t_{\rm eff}
    \left(
     \frac{\sigma_x}{t_{\rm MXM}}+\frac{\sigma_y}{t_{\rm MM}}
    \right)
   \biggr]
   {\mit\Phi}_s(x)
   {\rm d}x\,,
   \label{E:HACPc}
\end{eqnarray}
with the Pauli matrices $(\sigma_x,\sigma_y,\sigma_z)$
and a spinor notation
${\mit\Phi}_s^\dagger(x)=(\psi_s^{(r)}(x)^*,\psi_s^{(l)}(x)^*)$.
Here the Fermi velocity $v_{\rm F}$ and the dimensionless coupling
constant $\lambda$ has, respectively, been defined as
\begin{equation}
   \hbar v_{\rm F}=2at_{\rm eff}\,,
\end{equation}
\begin{equation}
   \lambda
   =\left\{
     \begin{array}{l}
      \beta_{\rm eff}^2/
       \pi t_{\rm eff}K_{\rm MX}\ \ \ 
      {\rm for}\ {\cal H}_{\rm eff}^{\rm CDW}\,,\\
      4\alpha_{\rm eff}^2/
       \pi t_{\rm eff}K_{\rm MX}\ \ \ 
      {\rm for}\ {\cal H}_{\rm eff}^{\rm ACP}\,.\\
     \end{array}
    \right.
\end{equation}

   Interestingly, both Hamiltonians (\ref{E:HCDWc}) and
(\ref{E:HACPc}) are mathematically equivalent to the TLM model
\begin{eqnarray}
   &&
   {\cal H}_{\rm TLM}
  =\int
   \frac{{\mit\Delta}(x)^2}{2\pi\hbar v_{\rm F}\lambda}
   {\rm d}x
  +\sum_s\int
   {\mit\Phi}_s^\dagger(x)
   \nonumber \\
   &&\quad
  \times
   \left[
    -{\rm i}\hbar v_{\rm F}\sigma_z\frac{{\rm d}}{{\rm d}x}
    +{\mit\Delta}(x)\sigma_x
   \right]
   {\mit\Phi}_s(x)
   {\rm d}x\,,
   \label{E:HTLM}
\end{eqnarray}
because they can exactly be transformed into Eq. (\ref{E:HTLM}),
except for the minus sign attached to the $\sigma_z$ term, by
rotating the spinor wave function around the $z$ axis as
${\rm exp}({\rm i}\varphi\sigma_z/2){\mit\Phi}_s(x)$ with
${\rm tan}\varphi$ being equal to $t_{\rm MM}/t_{\rm MXM}$ and
$t_{\rm MXM}/t_{\rm MM}$ for ${\cal H}_{\rm eff}^{\rm CDW}$ and
${\cal H}_{\rm eff}^{\rm ACP}$, respectively.
The Hamiltonian (\ref{E:HTLM}) possesses, regardless of the sign of
its $\sigma_z$ term, an exact solution of the form (\ref{E:WF})
\cite{T2388,K4173}.
Consequently the equivalent Hamiltonians (\ref{E:HCDWc}) and
(\ref{E:HACPc}) also give the soliton solution
\begin{equation}
   {\mit\Delta}(x)
   ={\mit\Delta}_0
    {\rm tanh}
    \left[
     (x-x_0)/\xi_{\rm s}
    \right]\,,
\end{equation}
with the TLM scaling relation and formation energy
\begin{equation}
   \xi_{\rm s}
   =\hbar v_{\rm F}/{\mit\Delta}_0
   =4at_{\rm eff}/E_{\rm g}\,,
   \label{E:xiTLM}
\end{equation}
\begin{equation}
   E_{\rm s}=2{\mit\Delta}_0/\pi=E_{\rm g}/\pi\,.
   \label{E:EsTLM}
\end{equation}
When we substitute the condition $t_{\rm MM}=2t_{\rm MXM}$ into
Eq. (\ref{E:xiTLM}), we obtain $\xi_{\rm s}=2a/\sqrt{5}E_{\rm g}$,
which explains well the numerical findings (\ref{E:xi}).
Figure \ref{F:Es} demonstrates the unique behavior of the solitons in
the weak-coupling region claiming that
\begin{equation}
   E_{\rm s}\simeq 0.32E_{\rm g}\,,
   \label{E:Es}
\end{equation}
which is also in good agreement with Eq. (\ref{E:EsTLM}).

\subsection{The Effective Mass}\label{SS:EM}

   The soliton mass $M_{\rm s}$ can be calculated by considering a
slowly-moving domain wall, that is, substituting $x_0=v_{\rm s}t$
into Eq. (\ref{E:WF}).
Neglecting any change in the wall shape, which must be of order
$v_{\rm s}^2$, we obtain
\begin{equation}
   \frac{1}{2}M_{\rm s}v_{\rm s}^2
  =\frac{1}{2}M
   \sum_n\left[
          \frac{{\rm d}}{{\rm d}t}(u_n-v_n)
         \right]^2
  =\frac{2Ml_0^2v_{\rm s}^2}{3a\xi}\,,
   \label{E:Ms}
\end{equation}
within the adiabatic approximation, where $M$ corresponds to the
halogen mass for R$_4$[Pt$_2$(pop)$_4$X]$\cdot$$n$H$_2$O with the CDW
ground state and to the mass of the binuclear platinum complex
Pt$_2$(dta)$_4$ for Pt$_2$(dta)$_4$I with the ACP ground state.

   Since there is little information available on electronic
parameters such as hopping amplitudes in MMX chains, we continue to
work with the condition $t_{\rm MM}=2t_{\rm MXM}$.
Among plenty of adjustable parameters, we lay great importance on the
band gap $E_{\rm g}$, which can reliably be measured in general.
We first discuss the pop complexes, which are similar to MX chains in
structure and therefore allow us to rely upon a good knowledge of the
MX system.
There is increased distortion of the halogen sublattice in the order
$\mbox{I}<\mbox{Br}<\mbox{Cl}$ \cite{B1155}, in which the halogen
mass oppositely decreases.
Considering their contributions to the soliton mass,
$M_{\rm s}\propto Ml_0^2$, solitons turn out most mobile in the iodo
complexes \cite{O250}.
The smaller gap, the larger mobility of solitons.
Hence we take a sample material
Na$_4$[Pt$_2$(pop)$_4$I]$\cdot$$7$H$_2$O \cite{M} with the CDW ground
state of $E_{\rm g}=1.2\,\mbox{eV}$.
Its structural data are available as $a=8.9\,\mbox{\AA}$ and
$l_0=0.3\,\mbox{\AA}$.
For lack of a few more parameters, we depend on calculations of PtI
compounds.
Baeriswyl and Bishop (BB) \cite{B239} estimate that
$t_{\rm MXM}=0.8\,\mbox{eV}$ and
$K_{\rm MX}=8.9\,\mbox{eV\AA}^{-2}$,
while Tagawa and Suzuki (TS) \cite{T2212} claim that
$t_{\rm MXM}=1.52\,\mbox{eV}$ and
$K_{\rm MX}=11.47\,\mbox{eV\AA}^{-2}$.
Standing on the experimental findings $E_{\rm g}=1.2\,\mbox{eV}$ and
$a=8.9\,\mbox{\AA}$ with the help of the BB and TS parametrizations,
the lattice distortion $l_0$ turns out $0.18\,\mbox{\AA}$ and
$0.13\,\mbox{\AA}$, respectively, the former of which is closer to
the observations.
The resultant soliton features are summarized in Table \ref{T:Mspop},
where $m_{\rm e}$ is the free-electron mass.
In any case the estimated mass is smaller than that for PtI chains,
$M_{\rm s}\sim 200m_{\rm e}$ \cite{O250}.
Polynuclear metal units contribute toward suppressing $l_0$ and
enlarging $a$ and thus end up with smaller soliton masses.
However, the soliton masses in the pop complexes still stay much
larger than those in polyacetylene, $M_{\rm s}\sim 6m_{\rm e}$
\cite{S1698,S2099}.
This is because most of the existent pop-family compounds lie in a
rather strong-coupling region.
Recently there has appeared an interesting attempt \cite{Y2321} to
tune the Peierls gap of the pop complexes by replacing their counter
ions.
Such explorations may further enhance the soliton mobility.

   Another possibility may lie in the dta complex.
Pt$_2$(dta)$_4$I exhibits a smaller gap \cite{K10068} probably due to
its smaller $\mbox{Pt}-\mbox{Pt}$ distance \cite{B444} and increased
mixing of halide character in the $d_\sigma$ $\mbox{Pt}-\mbox{Pt}$
wave function \cite{Y125124}.
The crystalline structure of Pt$_2$(dta)$_4$I is much different from
those of MX chains and therefore we do not have any useful
information on its electronic parameters.
Relying upon the experimental findings, $E_{\rm g}=0.3\,\mbox{eV}$
\cite{K10068} and $a=8.6\,\mbox{\AA}$ \cite{B444}, and assuming that
$t_{\rm MM}=2t_{\rm MXM}=2\,\mbox{eV}$ and $1.5\,\mbox{eV}$ together
with $K_{\rm MX}=8.9\,\mbox{eV\AA}^{-2}$, we give trial estimates in
Table \ref{T:Msdta}.
Considering the smaller $\mbox{Pt}-\mbox{Pt}$ distance, the results
with $t_{\rm MM}=1.5\,\mbox{eV}$ may be too conservative.
Anyway Pt$_2$(dta)$_4$I is likely to have much more mobile solitons
than the pop complexes.

   Metal polynucleation leads to a smaller gap and thus suppresses
the soliton mass.
This must be the case with the pop complexes, more generally, with
the family compounds exhibiting the ground states of the CDW type.
However, it may not necessarily be the case with the dta complex,
that is, with the family compounds which possess neutral chain
structures and therefore exhibit the ACP-type ground states, where
increasing $M$ may prevent $M_{\rm s}$ from decreasing.
The usual metallic conduction, rather than a solitonic one, may be
much more relevant in such polynuclear metal complexes.

\section{Summary}\label{S:S}

   We have demonstrated the soliton excitations in MMX chains laying
emphasis on both similarities and differences between the cases of
the CDW and ACP backgrounds.
Solitons in the weak-coupling region are all degenerate to each other
and are uniquely scaled by the band gap, whether it originates in
the CDW or ACP instability.
The degeneracy is lifted in the strong-coupling region, where neutral
solitons come to have higher energies than charged solitons.
Solitons  are likely to be more mobile in MMX chains than in MX
chains.
We are hoping for a systematic observation on various binuclear and
possibly polynuclear \cite{S8366} metal complexes.

\acknowledgments

   The authors are grateful to Prof. K. Yonemitsu for fruitful
discussion.
They further thank Prof. H. Okamoto for helpful information on MMX
materials.
This work was supported by the Japanese Ministry of Education,
Science, Sports and Culture.
The numerical calculation was done using the facility of the
Supercomputer Center, Institute for Solid State Physics, University
of Tokyo.

\begin{table}
\caption{Two estimates of the soliton width and mass fitting
         Na$_4$[Pt$_2$(pop)$_4$I]$\cdot$$7$H$_2$O of
         $E_{\rm g}=1.2\,\mbox{eV}$ and $a=8.9\,\mbox{\AA}$, where
         we assume the BB and TS parametrizations (see text).}
\begin{tabular}{lcccc}
&
\multicolumn{2}{c}{\ BB parameters\ } &
\multicolumn{2}{c}{\ TS parameters\ } \\
\cline{2-3}
\cline{4-5}
&
\ $\,\xi_{\rm s}/a$ & $M_{\rm s}/m_{\rm e}$ & 
\ $\,\xi_{\rm s}/a$ & $M_{\rm s}/m_{\rm e}$ \\
\hline
\ S$^-$ & \ $\,1.2$ & $ 94$ & \ \ $2.4$ & $28$ \\
\ S$^+$ & \ $\,0.7$ & $160$ & \ \ $1.6$ & $42$ \\
\ S$^0$ & \ $\,1.1$ & $100$ & \ \ $2.1$ & $31$ \\
\end{tabular}
\label{T:Mspop}
\end{table}

\begin{table}
\caption{Two estimates of the soliton width and mass fitting
         Pt$_2$(dta)$_4$I of $E_{\rm g}=0.3\,\mbox{eV}$ and
         $a=8.6\,\mbox{\AA}$, where we assume
         $t_{\rm MM}=2.0\,\mbox{eV}$ and
         $t_{\rm MM}=1.5\,\mbox{eV}$.}
\begin{tabular}{lcccc}
&
\multicolumn{2}{c}{\ \ $t_{\rm MM}=2.0\,\mbox{eV}$\ \ } &
\multicolumn{2}{c}{\ \ $t_{\rm MM}=1.5\,\mbox{eV}$\ \ } \\
\cline{2-3}
\cline{4-5}
&
\ $\,\xi_{\rm s}/a$ & $M_{\rm s}/m_{\rm e}$ & 
\ $\,\xi_{\rm s}/a$ & $M_{\rm s}/m_{\rm e}$ \\
\hline
\ S$^-$ & \ $\,6.2$ & $12$ & \ \ $4.7$ & $19$ \\
\ S$^+$ & \ $\,6.0$ & $12$ & \ \ $4.5$ & $20$ \\
\ S$^0$ & \ $\,6.1$ & $12$ & \ \ $4.6$ & $19$ \\
\end{tabular}
\label{T:Msdta}
\end{table}

\widetext
\end{document}